\documentclass[conference]{IEEEtran}

\usepackage{cite}
\usepackage{amsmath,amssymb,amsfonts}
\usepackage{algorithmic}
\usepackage{graphicx}
\usepackage{textcomp}
\usepackage{xcolor}
\usepackage{booktabs}
\usepackage{siunitx}
\usepackage{hyperref}
\usepackage{url}
\usepackage{balance}
\usepackage{tikz}
\usetikzlibrary{arrows.meta, shapes.geometric, positioning, calc}

\graphicspath{{fig/}}

\begin{document}
\bstctlcite{BSTcontrol}

\title{From Least Squares to Deep Learning: Benchmarking Indoor Positioning on the HYMN Multi-Technology Dataset}

\author{
\IEEEauthorblockN{Paul Schwarzbach and Muhammad Ammad}
\IEEEauthorblockA{Chair of Transport Systems Information Technology, TUD Dresden University of Technology, Germany}
}

\maketitle

\begin{abstract}
Indoor positioning benchmarks rarely co-locate multiple technologies with high-accuracy ground truth, limiting cross-method and cross-technology comparison. We evaluate four range-based positioning methods on fused Ultra-wideband, Bluetooth Low Energy, and WiFi ranges recorded at 48 reference points in an industrial facility with available ground truth. The evaluated methods range from least-squares positioning, with and without robust weighting, through a Bayesian grid filter to a ResNet regressor that we report under two protocols, interpolation and spatial generalisation. Per-technology ranging quality spans roughly two orders of magnitude, which caps any geometry-based solver that weighs anchors equivalently. On fused input the grid filter is competitive with deep learning on median error, while the ResNet's advantage concentrates on the upper tail. Holding reference points out of training increases the learned regressor's median error several-fold, exposing a spatial-generalisation penalty invisible under random splits. The dataset and evaluation code are publicly available.
\end{abstract}

\section{Introduction}
\label{sec:introduction}

Reliable indoor positioning has become a prerequisite for applications ranging from warehouse logistics and industrial asset tracking to emergency response and smart building management.
Ultra-wideband (UWB) offers centimeter-level ranging but limited range, while Bluetooth Low Energy (BLE) and WiFi provide broad coverage at coarser resolution. Fifth-generation (5G) cellular signals add wide-area reach, and Global Navigation Satellite Systems (GNSS) remain the default outdoors yet degrade indoors.
Multiple surveys document these technologies and their positioning principles~\cite{Zafari2019_survey, Yassin2017_advances, Laoudias2018_enabling, Li2021_LEIoT}.

Most positioning studies evaluate a single radio technology or a single algorithm family, which limits the generalizability of reported results~\cite{MendozaSilva2019_metareview}.
Public benchmark datasets that combine multiple co-located radio technologies with sub-centimeter ground truth are scarce. Widely used collections such as UJIndoorLoc~\cite{uji} or the Tampere dataset~\cite{TampereU1} cover WiFi fingerprints only, and a systematic review of open research data in indoor positioning confirms that multi-technology datasets are underrepresented~\cite{ordip, Ayub2025_Indoor_Datasets_study}.
At the same time, the gap between classical geometric positioning methods and data-driven deep learning approaches is seldom quantified on the same measurement data, so practitioners lack clear guidance on when added model complexity pays off.
Reported results are also rarely independently verifiable, because the measurements, processing code, and trained-model recipes behind a single error figure are seldom released together.

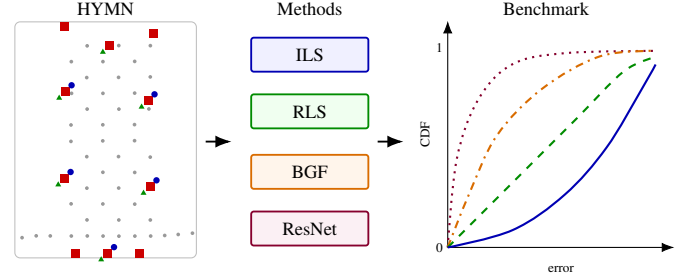
\begin{figure}[!th]
  \centering
  \begin{tikzpicture}[
    font=\scriptsize,
    >=Latex,
    panel title/.style={font=\scriptsize, anchor=south, inner sep=1pt},
    methodbox/.style={draw, line width=0.7pt, rounded corners=1pt, minimum width=1.55cm, minimum height=0.45cm, font=\scriptsize, align=center, inner sep=1pt},
    flow/.style={->, line width=0.7pt, shorten >=1.5pt, shorten <=1.5pt}
  ]

    %% =========================================================
    %% Panel (a): HYMN floor plan with real coordinates
    %% X_LOCAL range approx [-20.2, 12.1] m;  Y_LOCAL range approx [-6.4, 36.6] m
    %% xscale = yscale = 0.07 cm/m preserves aspect
    %% =========================================================
    \begin{scope}[local bounding box=panelA, xscale=0.07, yscale=0.07]
      % Bounding rectangle around the surveyed region (slight padding)
      \draw[thin, gray!50, rounded corners=2pt] (-21.5, -7.2) rectangle (12.8, 37.4);

      % 48 reference points (X_LOCAL_CENTER, Y_LOCAL_CENTER from point_coordinates.csv)
      \foreach \x/\y in {%
        4.104/32.903, 1.141/29.928, -1.866/32.92, -4.86/29.889, -7.839/32.869,
        -10.831/29.882, -7.922/26.872, -10.792/23.891, -4.86/23.891, -1.907/26.909,
        1.136/23.935, 4.089/26.995, 3.827/20.947, 1.109/17.981, -1.841/20.912,
        -4.884/17.95, -7.842/20.916, -10.893/17.915, 4.083/11.839, 1.146/8.945,
        -1.916/11.94, -4.87/8.975, -7.915/11.927, -10.875/9.058, -10.872/3.01,
        -7.895/5.977, -4.875/2.989, -1.882/5.995, 1.149/2.955, 4.155/5.798,
        4.096/-0.04, -1.859/-0.04, -7.871/-0.026, 1.118/14.944, -4.892/14.912,
        -10.902/14.812, -20.14/-3.382, -17.467/-3.033, -14.809/-3.049,
        -10.833/-3.07, -7.814/-3.057, -4.836/-3.078, -1.841/-3.097, 1.175/-3.083,
        4.162/-3.074, 6.813/-2.964, 9.365/-2.779, 12.056/-2.534%
      }{
        \node[circle, fill=black!40, inner sep=0pt, minimum size=1.4pt] at (\x,\y) {};
      }

      % UWB anchors (10) — red squares, plotted first so others can sit on top
      \foreach \x/\y in {%
        3.685/22.496, 4.059/6.256, -3.969/-6.381, -12.157/7.832, -11.976/24.21,
        -3.651/32.95, 4.754/35.133, 2.035/-6.415, -9.984/-6.38, -12.099/36.584%
      }{
        \node[rectangle, fill=red!80!black, inner sep=0pt, minimum size=3.2pt] at (\x,\y) {};
      }

      % BLE anchors (5) — blue circles, slight up-right offset
      \foreach \x/\y in {%
        3.69/22.509, 4.058/6.255, -3.964/-6.37, -12.172/7.83, -11.984/24.218%
      }{
        \node[circle, fill=blue!70!black, inner sep=0pt, minimum size=2.4pt] at ($(\x,\y)+(1.2,1.2)$) {};
      }

      % WiFi anchors (6) — green triangles, slight down-left offset
      \foreach \x/\y in {%
        3.644/22.526, 4.012/6.247, -3.988/-6.327, -12.104/7.831, -11.927/24.211,
        -3.673/32.893%
      }{
        \node[regular polygon, regular polygon sides=3, fill=green!55!black, inner sep=0pt, minimum size=2.6pt] at ($(\x,\y)+(-1.2,-1.2)$) {};
      }
    \end{scope}

    %% =========================================================
    %% Arrow (a) -> (b)
    %% =========================================================
    \draw[flow] ([xshift=2pt]panelA.east) -- ++(0.45, 0);

    %% =========================================================
    %% Panel (b): four method boxes stacked vertically, anchored to panelA top
    %% =========================================================
    \begin{scope}[shift={($(panelA.east)+(1.5, 0)$)}, local bounding box=panelB]
      \node[methodbox, draw=blue!70!black,   fill=blue!6]   (b-ils) at (0,  1.15) {ILS};
      \node[methodbox, draw=green!55!black,  fill=green!6]  (b-rls) at (0,  0.40) {RLS};
      \node[methodbox, draw=orange!85!black, fill=orange!8] (b-bgf) at (0, -0.40) {BGF};
      \node[methodbox, draw=purple!70!black, fill=purple!6] (b-res) at (0, -1.15) {ResNet};
    \end{scope}

    %% =========================================================
    %% Arrow (b) -> (c)
    %% =========================================================
    \draw[flow] ([xshift=2pt]panelB.east) -- ++(0.45, 0);

    %% =========================================================
    %% Panel (c): teaser ECDF; axes span 1.8 x 1.8 cm
    %% Curve order matches observed: ResNet < BGF < RLS < ILS
    %% =========================================================
    %% Panel (c) origin sits at the axis cross; axes use local coords starting at (0,0)
    \begin{scope}[shift={($(panelB.east)+(1.05, -1.4)$)}, local bounding box=panelC]
      % Axes: x in [0, 2.9], y in [0, 2.9] -> 2.9 x 2.9 cm
      \draw[->, thin] (0, 0) -- (3.0, 0);
      \draw[->, thin] (0, 0) -- (0, 3.0);
      \node[font=\tiny, inner sep=1pt] at (1.5, -0.28) {error};
      \node[font=\tiny, rotate=90, inner sep=1pt] at (-0.3, 1.5) {CDF};
      \node[font=\tiny, inner sep=0.5pt, anchor=east] at (-0.04, 2.65) {1};
      \node[font=\tiny, inner sep=0.5pt, anchor=east] at (-0.04, 0.0) {0};

      \draw[purple!70!black, thick, dotted]
        plot[smooth, tension=0.6] coordinates {(0,0) (0.12,1.17) (0.40,1.95) (0.85,2.42) (1.55,2.57) (2.75,2.60)};
      \draw[orange!85!black, thick, dash dot]
        plot[smooth, tension=0.6] coordinates {(0,0) (0.28,0.70) (0.65,1.48) (1.25,2.10) (2.02,2.53) (2.75,2.60)};
      \draw[green!55!black, thick, dashed]
        plot[smooth, tension=0.6] coordinates {(0,0) (0.47,0.47) (1.09,1.09) (1.71,1.71) (2.33,2.33) (2.75,2.52)};
      \draw[blue!70!black, thick]
        plot[smooth, tension=0.6] coordinates {(0,0) (0.86,0.23) (1.55,0.70) (2.10,1.32) (2.57,2.10) (2.75,2.42)};
    \end{scope}

    %% =========================================================
    %% Panel titles — all placed at a common y (just above panelA.north)
    %% =========================================================
    \node[panel title] (titleA) at ([yshift=0.05cm]panelA.north) {HYMN};
    \node[panel title] at (panelB.north |- titleA.south) {Methods};
    \node[panel title] at (panelC.north |- titleA.south) {Benchmark};

  \end{tikzpicture}
  \caption{Benchmarking four families of indoor positioning methods on the multi-technology HYMN dataset~\cite{HYMN_Dataset_zenodo_2025}.}
  \label{fig:graphical_abstract}
\end{figure}

The openly available HYMN (Hybrid Multi-technology Navigation) dataset~\cite{HYMN_Dataset_zenodo_2025, Ammad2026_HYMN_descriptor} provides time-synchronised measurements from UWB, BLE, WiFi, 5G, and GNSS, collected at 48 reference points in an industrial facility with millimetre-level ground truth. We apply iterative and robust least squares (ILS, RLS), a Bayesian grid filter (BGF), and a ResNet regressor to the fused multi-technology positioning solution, and we evaluate the ResNet under both a random-split and a reference-point-held-out protocol to measure spatial generalisation. The fused positioning solution is built from the three ranging technologies (UWB, BLE, WiFi). 5G and indoor GNSS are not used as fusion inputs. Per-technology ranging quality is reported here as evaluation context. A deeper ranging evaluation of the indoor/outdoor transition zone of the dataset, which also extends to GNSS, is given by Michler et al.~\cite{Michler2025}, with a cross-technology ranging-residual analysis indoors in~\cite{Schwarzbach2026_Bridging_the_Gap}.

This paper makes three contributions.
\begin{enumerate}
    \item We benchmark ILS, RLS, BGF, and a ResNet regressor on the multi-technology HYMN dataset under a common range-based measurement model, reporting their performance on the fused positioning solution.
    \item For the ResNet, we contrast interpolation and spatial-generalisation performance across a random-split and a reference-point-held-out protocol, quantifying a generalisation gap that random-split benchmarks do not surface.
    \item We contribute an open, end-to-end reproducible benchmark, releasing the HYMN measurements, the evaluation pipeline, and the trained-model recipes under MIT licenses with persistent DOIs.
\end{enumerate}

%Section~\ref{sec:related_work} reviews positioning method families and surveys public multi-technology benchmark datasets. Section~\ref{sec:methodology} presents the measurement model and the four evaluated methods. Section~\ref{sec:evaluation} reports the ranging and positioning results with integrated discussion. Section~\ref{sec:conclusion} concludes.

\section{Related Work}
\label{sec:related_work}

Indoor positioning methods can be grouped along two axes.
The first is measurement type: received signal strength (RSS), time of arrival (ToA), time difference of arrival (TDoA), or angle of arrival (AoA).
The second is the solution family, distinguishing geometric approaches that exploit known signal propagation models from data-driven approaches that learn a mapping from radio observations to position.
Several surveys offer detailed taxonomies of these approaches~\cite{Zafari2019_survey, MendozaSilva2019_metareview, yang2024positioningusingwirelessnetworks}, and dedicated reviews cover machine learning~\cite{Nessa2020} and deep learning~\cite{Kordi2024_dlsurvey} for indoor positioning.
The four methods evaluated in this paper span both families: iterative and robust least squares represent the geometric branch, the Bayesian grid filter operates on a probabilistic state space, and the ResNet regressor is a purely data-driven approach.

\paragraph{Geometric positioning}
Range-based positioning solves a set of nonlinear distance equations by least squares (LS).
Foy~\cite{FOY1976_Taylor_LS} introduced the iterative Gauss--Newton scheme (ILS) that linearises the distance equations around an initial estimate, and robust variants down-weight outliers caused by non-line-of-sight (NLOS) propagation~\cite{Yassin2017_advances, Vaghefi2013}.
These methods are lightweight and need no training data, but they require known anchor positions and degrade when a substantial fraction of measurements is NLOS-corrupted. Data-driven methods avoid both requirements, yet risk degrading when deployment conditions differ from those seen during training.

\paragraph{Bayesian and probabilistic positioning}
Bayesian filters maintain a posterior over the position state and update it with each measurement.
Grid-based (histogram) filters discretise the state space into cells, as applied to robot localisation by Burgard~et~al.~\cite{Burgard_probability_grids_1996} and formalised by Thrun~et~al.~\cite{Thrun2005}.
Particle filters offer a continuous-state alternative with adaptive resolution~\cite{Fox2003}.
In the HYMN setting, the constant-velocity motion update, tuned for near-static reference points, smooths the belief between epochs, suppressing short-lived outliers and letting the filter recover from a poor initial estimate as evidence accumulates.

\paragraph{Deep learning for indoor positioning}
Deep learning enters indoor positioning either as direct regression from radio features such as RSS, channel state information (CSI), and the channel impulse response to coordinates, or as NLOS identification and ranging-error mitigation upstream of a geometric solver.
Surveys~\cite{Kordi2024_dlsurvey, Sonny2024_mlwireless} catalogue convolutional, recurrent, and attention-based architectures.
Representative applications span WiFi fingerprinting~\cite{Kim2018_scalableDNN}, UWB coordinate regression~\cite{Poulose2020_uwblstm}, UWB range-error mitigation~\cite{Stahlke2021_edgedl}, and CSI-based localisation~\cite{Zhang2023_csiattention} built on the ResNet backbone of He~et~al.~\cite{He2016_resnet}.
A common limitation is the focus on a single radio technology.
Our prior work~\cite{Ammad2025_Rangle_Angle_Resnet} introduced range-angle likelihood maps as input to a ResNet for UWB aircraft cabin positioning, which this study extends to multi-technology input on a different dataset.

\paragraph{Multi-technology fusion and benchmark datasets}
Multi-technology positioning has been explored for specific technology pairs.
Bai~et~al.~\cite{Bai_gnss_5g_hybrid_multi_rate_measurements_2022} fused GNSS and 5G measurements in an extended Kalman filter, and Leitch~et~al.~\cite{Leitch2023} combined WiFi, BLE, UWB, and inertial sensors for indoor localization.
Michler~et~al.~\cite{Michler2025} assessed signals of opportunity from multiple systems for indoor-outdoor transitions.
To our knowledge, no prior benchmark jointly evaluates several positioning method families on a shared multi-technology indoor dataset of this scope.
On the dataset side, publicly available indoor positioning datasets are dominated by single-technology WiFi fingerprint collections~\cite{ordip, uji, TampereU1}.
Recent multi-modal contributions such as the WiFi--CCTV dataset of Abdalla~et~al.~\cite{Abdalla2026_WiFi_CCTV_Dataset} broaden the scope but do not include ranging technologies like UWB.
The HYMN dataset~\cite{HYMN_Dataset_zenodo_2025, Ammad2026_HYMN_descriptor} addresses this gap. Section~\ref{subsec:dataset} describes its measurements, and the companion ranging analysis is reported in~\cite{Michler2025, Schwarzbach2026_Bridging_the_Gap}.

\section{Methodology}
\label{sec:methodology}

We evaluate four methods spanning the solution families used for range-based positioning: iterative and robust least squares (ILS, RLS), a Bayesian grid filter (BGF), and a ResNet regressor. They share the measurement model below and differ in how they treat heavy-tailed errors and whether they need training data. Fingerprinting regressors such as k-nearest neighbours instead operate on dense radio-map surveys, a different observation modality outside this comparison, and are left to future work.
Derivations for the classical methods and the ResNet architecture are given in our prior works~\cite{Schwarzbach2026_ResNet, Ammad2025_Rangle_Angle_Resnet}. The descriptions below focus on the formulations and hyperparameters used for the HYMN evaluation.

Let $\mathbf{a}_j \in \mathbb{R}^2$ denote the known position of anchor~$j$, $r_{ij}$ the measured range from that anchor at epoch~$i$, and $\hat{\mathbf{p}}_i \in \mathbb{R}^2$ the position estimate for that epoch with reference $\mathbf{p}_i^{\text{ref}}$.
At each epoch, the BLE, UWB, and WiFi anchor ranges available at the tag are concatenated into a single set $\{(\mathbf{a}_j, r_{ij})\}_{j=1}^{N_i}$, where $N_i$ is the number of anchors that provide a range at epoch~$i$. We retain only epochs with $N_i \geq 3$ anchors.
The optimisation-based methods share the residual $e_{ij}(\mathbf{p}) = r_{ij} - \|\mathbf{p} - \mathbf{a}_j\|_2$, the epoch-$i$ difference between the measured range to anchor~$j$ and the range at a candidate position~$\mathbf{p}$, and differ in the loss $\rho$ applied to it,
\begin{equation}
  \hat{\mathbf{p}}_i = \arg\min_{\mathbf{p}} \sum_{j=1}^{N_i} \rho\!\left( e_{ij}(\mathbf{p}) \right).
  \label{eq:generic_cost}
\end{equation}
The loss $\rho$ for ILS and RLS, and the likelihood for BGF, are defined in the subsections below.
The ResNet, in contrast, learns a direct mapping from a likelihood-map representation of the same ranges to the position, without solving~\eqref{eq:generic_cost} at inference time. It thus front-loads its cost into offline training and reduces inference to a single forward pass, unlike the per-epoch iterative solve of the other methods.

\subsection{Dataset}
\label{subsec:dataset}

The HYMN dataset~\cite{HYMN_Dataset_zenodo_2025, Ammad2026_HYMN_descriptor} provides time-synchronised measurements from five positioning technologies in an industrial hall in Torgau, Germany, measuring $44 \times 18$~m.
Gates on each side of the hall were open throughout the campaign, placing part of the measurement grid in a transition zone with mixed indoor and outdoor propagation.
Ground truth for every reference point and anchor position is surveyed with a Leica TS16 total station at millimeter precision and expressed in a common local frame.

The installed infrastructure relevant for this work includes:
\begin{itemize}
  \item \textbf{UWB}: 10 two-way ranging (TWR) anchors.
  \item \textbf{BLE}: 5 phase-based ranging (PBR) beacons.
  \item \textbf{WiFi}: 6 fine-timing-measurement (FTM) access points.
\end{itemize}

Data is collected at 48 reference points (36 indoor, 12 in the transition zone), with three minutes of static recording per point.
The raw sample counts exceed 60k per ranging technology.
Ranging-quality analyses for the same measurement campaign are reported in companion works~\cite{Michler2025, Schwarzbach2026_Bridging_the_Gap}.

\subsection{Iterative Least Squares (ILS)}
\label{subsec:ils}

ILS solves~\eqref{eq:generic_cost} with the quadratic loss $\rho(e) = e^2$ via Gauss--Newton iteration~\cite{FOY1976_Taylor_LS, Norrdine2012, Vaghefi2013}.
Linearising the residual around the current estimate $\mathbf{p}^{(k)}$ and stacking the per-anchor rows $\partial e_{ij}/\partial \mathbf{p}$ into the Jacobian $\mathbf{J}^{(k)} \in \mathbb{R}^{N_i \times 2}$ gives the update
\begin{equation}
  \mathbf{p}^{(k+1)} = \mathbf{p}^{(k)} - \big(\mathbf{J}^{(k)\top}\mathbf{J}^{(k)}\big)^{-1} \mathbf{J}^{(k)\top} \mathbf{e}^{(k)},
  \label{eq:gauss_newton}
\end{equation}
which we solve via a singular-value decomposition to stay numerically stable when the anchor geometry becomes near-planar.
The iteration is initialised at the anchor centroid, runs at most seven steps, and terminates when $\|\mathbf{p}^{(k+1)} - \mathbf{p}^{(k)}\|_2 < 10^{-6}$.
ILS is inexpensive and yields a closed-form covariance, but quadratic weighting lets one grossly corrupted range dominate the sum and bias the solution.

\subsection{Robust Least Squares (RLS)}
\label{subsec:rls}

RLS retains the residual of ILS but replaces the quadratic loss with the Huber function~\cite{Huber1964},
\begin{equation}
  \rho_H(e) = \begin{cases}
    \tfrac{1}{2} e^2 & |e| \leq \delta, \\
    \delta\, |e| - \tfrac{1}{2} \delta^2 & |e| > \delta,
  \end{cases}
  \label{eq:huber}
\end{equation}
so that the influence of a residual grows linearly, rather than quadratically, once it exceeds the threshold~$\delta$.
We set $\delta = \SI{1.0}{\meter}$, reflecting the scale at which UWB residuals begin to deviate from their Gaussian bulk (Section~\ref{subsec:ranging}).
The minimisation uses the SciPy trust-region reflective algorithm, initialised at the anchor centroid and capped at $50\,N_i$ function evaluations.
RLS buys outlier tolerance over ILS at modest additional cost, but cannot recover the position when residuals on most anchors exceed $\delta$ by a large factor, a regime the BLE and WiFi ranging statistics approach (Section~\ref{subsec:ranging}).

\subsection{Bayesian Grid Filter (BGF)}
\label{subsec:bgf}

The BGF replaces point estimation with maintenance of a posterior over position~\cite{Thrun2005, Fox2003, Schwarzbach2020_tight_integration}.
We discretise the region bounded by the anchor hull plus a \SI{2}{\meter} padding into a regular grid $\mathcal{G}$ with cell size $\Delta g = \SI{0.5}{\meter}$ and initialise a uniform prior over $\mathcal{G}$.
Each range $r_{ij}$ contributes a Gaussian likelihood,
\begin{equation}
  \mathcal{L}_{ij}(\mathbf{g}) = \exp\!\left( -\frac{(r_{ij} - \|\mathbf{g} - \mathbf{a}_j\|_2)^2}{2\,\sigma_{\tau(j)}^2} \right),
  \label{eq:bgf_likelihood}
\end{equation}
where $\tau(j)$ denotes the technology of anchor~$j$ and $(\sigma_{\text{UWB}}, \sigma_{\text{BLE}}, \sigma_{\text{WiFi}}) = (\SI{0.27}{\meter}, \SI{4.76}{\meter}, \SI{1.46}{\meter})$ are set to the per-technology median absolute ranging residuals reported in Section~\ref{subsec:ranging}.
Between epochs the belief is diffused by a constant-velocity random-walk: the predicted displacement has mean $\bar v\,\Delta t$ and standard deviation $\sigma_v\,\Delta t$, with $\bar v = \SI{0.1}{\meter\per\second}$, $\sigma_v = \SI{0.2}{\meter\per\second}$, and $\Delta t$ the epoch-to-epoch timestamp difference floored at \SI{0.2}{\second}.
The posterior at epoch~$i$ is the prediction multiplied by the joint range likelihood and normalised,
\begin{equation}
  p_i(\mathbf{g}) \propto \bar p_i(\mathbf{g}) \prod_{j=1}^{N_i} \mathcal{L}_{ij}(\mathbf{g}).
  \label{eq:bgf_posterior}
\end{equation}
The estimate $\hat{\mathbf{p}}_i$ is the probability-weighted mean of all cells within a \SI{2}{\meter} radius of the maximum-a-posteriori cell. Restricting the mean to this radius suppresses spurious secondary modes while still recovering the peak location when the posterior mass concentrates on a single mode.

\subsection{ResNet Regressor}
\label{subsec:resnet}

The ResNet is a supervised deep-learning regressor that maps a per-anchor likelihood-map tensor to coordinates, extending the residual architecture validated in~\cite{Ammad2025_Rangle_Angle_Resnet, Schwarzbach2026_ResNet} to the multi-technology HYMN input.
At each epoch the per-anchor Gaussian likelihoods of~\eqref{eq:bgf_likelihood} are evaluated on $\mathcal{G}$ and stacked as a multi-channel input. One channel per anchor preserves geometry-specific information that the BGF collapses into a single joint likelihood.
A four-stage residual convolutional network with $64/128/256/512$ filters, batch normalisation, ReLU activation, and dropout~\cite{He2016_resnet} flattens into a fully connected output layer that regresses $(\hat{x}_i, \hat{y}_i)$.
Training minimises the mean-squared error loss
\begin{equation}
  \mathcal{L}_{\text{MSE}} = \frac{1}{B} \sum_{i=1}^{B} \|\hat{\mathbf{p}}_i - \mathbf{p}_i^{\text{ref}}\|_2^2
  \label{eq:mse}
\end{equation}
over a batch of $B$ labelled epochs with the Adam optimiser.

Fewer than 7\% of published indoor-positioning datasets document an explicit train/test split convention~\cite{ordip}, yet reproducible comparison across methods requires consistent, explicitly defined data splits, task definitions, and evaluation metrics~\cite{pan2026aipositioning}.
We therefore define two complementary partitioning protocols and report results for each separately.
\texttt{ResNet-RandomSplit} partitions epochs uniformly at random across all 48 reference points, so every point contributes to both training and test sets. The protocol isolates the best-case interpolation regime within the surveyed area.
\texttt{ResNet-SpatialHoldout} partitions by reference-point identity (leave-points-out), so the test epochs come from locations never seen during optimisation. The protocol quantifies spatial generalisation to unseen positions.
The two variants share architecture, optimiser, and likelihood-map input, and differ only in the train/test partitioning rule.
The architecture and hyperparameters are not tuned on HYMN but reuse the static configuration validated in our prior work~\cite{Ammad2025_Rangle_Angle_Resnet, Schwarzbach2026_ResNet}, so the test splits serve only for final evaluation and never for model selection. Both variants and the train/test partitioning run with a single fixed random seed, which only fixes the split and initialisation for exact reproducibility, and all hyperparameters are listed in \texttt{config.py} of the companion repository.

\section{Results and Discussion}
\label{sec:evaluation}

\subsection{Error Metric}
\label{subsec:metric}

We report the horizontal positioning error $e_i = \lVert \hat{\mathbf{p}}_i - \mathbf{p}_i^{\text{ref}} \rVert_2$, with $\hat{\mathbf{p}}_i = (\hat{x}_i, \hat{y}_i)^\top$ the estimate and $\mathbf{p}_i^{\text{ref}}$ the surveyed reference position. HYMN ground truth is two-dimensional, so no vertical component is included. For each $(\text{technology}, \text{method})$ pair we summarise the empirical distribution of $e_i$ by the count $n$, the mean, median, standard deviation, and the 75th and 95th percentiles (P75, P95). Here $n$ counts the evaluated epochs. Ranging performance in Section~\ref{subsec:ranging} follows the same convention on the absolute per-anchor residual $|r_{ij} - \lVert \mathbf{a}_j - \mathbf{p}_i^{\text{ref}} \rVert_2|$, where $r_{ij}$ is the measured range to anchor $\mathbf{a}_j$ and $n$ instead counts individual anchor measurements.

\subsection{Ranging Performance}
\label{subsec:ranging}

\begin{figure}[t]
  \centering
  \includegraphics[width=\columnwidth]{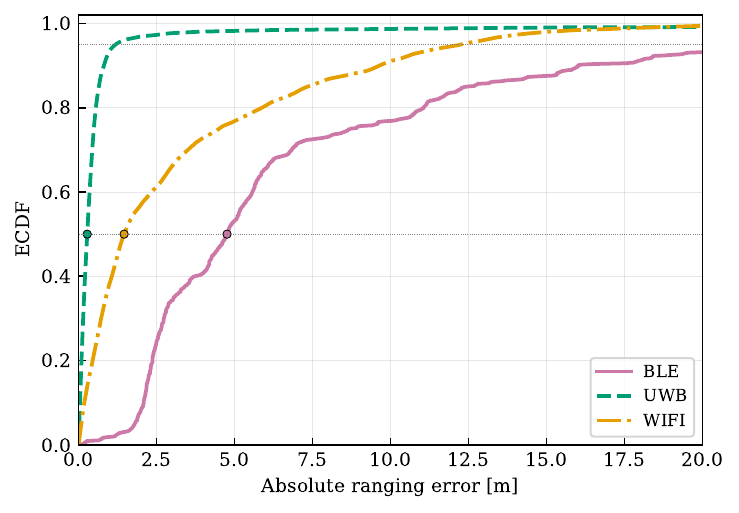}
  \caption{ECDF of absolute ranging residuals per technology. Circles mark the median. The abscissa is clipped at \SI{20}{\meter}.}
  \label{fig:ecdf_ranging}
\end{figure}

Figure~\ref{fig:ecdf_ranging} shows the empirical cumulative distribution function (ECDF) of the absolute ranging residual for BLE, UWB, and WiFi. UWB achieves a median residual of \SI{0.27}{\meter} with \SI{95}{\percent} of samples below \SI{1.18}{\meter} ($n = \num{62745}$). WiFi shows a median of \SI{1.46}{\meter} and a P95 of \SI{12.25}{\meter} ($n = \num{63121}$). BLE is widest, with a median of \SI{4.76}{\meter} and a P95 of \SI{26.77}{\meter} ($n = \num{49745}$).

The three technologies span roughly two orders of magnitude in median accuracy, which sets the ceiling for any geometry-based solver that treats all anchor ranges equivalently. UWB's mean (\SI{1.11}{\meter}) sits close to its P95 (\SI{1.18}{\meter}) despite a median of only \SI{0.27}{\meter}, because the top five percent of residuals carries extreme outliers that dominate the arithmetic mean of an otherwise tight distribution. A deeper characterisation of the ranging physics is given in \cite{Michler2025,Schwarzbach2026_Bridging_the_Gap}.

\subsection{Positioning Performance}
\label{subsec:positioning}

Table~\ref{tab:positioning_stats} reports the positioning error for the five evaluated methods on the fused multi-technology input. The classical methods (ILS, RLS, BGF) are computed on $n = \num{6075}$ measurement epochs. The two ResNet variants run on the hold-out splits defined in Section~\ref{sec:methodology} ($n = \num{1833}$ for \texttt{SpatialHoldout}, $n = \num{1823}$ for \texttt{RandomSplit}).

\begin{table}[t]
  \centering
  \caption{Summary of horizontal positioning error.}
  \label{tab:positioning_stats}
  \setlength{\tabcolsep}{4pt}
  \resizebox{\columnwidth}{!}{% auto-generated by evaluation/stats.py
\begin{tabular}{lrrrrrr}
\toprule
Method & $n$ & Mean [m] & Median [m] & Std [m] & P75 [m] & P95 [m] \\
\midrule
ILS & 6075 & 5.519 & 3.616 & 5.974 & 5.951 & 18.659 \\
RLS & 6075 & 2.013 & 1.075 & 3.744 & 1.930 & 5.684 \\
BGF & 6075 & 1.586 & \textbf{0.444} & 4.291 & \textbf{0.793} & 6.295 \\
ResNet-SpatialHoldout & 1833 & 3.504 & 3.222 & 1.915 & 4.786 & 7.565 \\
ResNet-RandomSplit & 1823 & \textbf{0.605} & 0.554 & \textbf{0.403} & 0.808 & \textbf{1.314} \\
\bottomrule
\end{tabular}
}
\end{table}

BGF attains the lowest median at \SI{0.44}{\meter} and the lowest P75 at \SI{0.79}{\meter}, but a small fraction of epochs collapse near the grid bounds and stretch its P95 to \SI{6.29}{\meter}. ResNet-RandomSplit sits within \SI{0.11}{\meter} of BGF at the median (\SI{0.55}{\meter}) and dominates the tail with a P95 of \SI{1.31}{\meter}. ILS retains the full outlier sensitivity of an unweighted nonlinear least-squares formulation, with a median of \SI{3.62}{\meter} and a P95 of \SI{18.66}{\meter}. The Huber-weighted RLS recovers a usable median of \SI{1.07}{\meter} with a P95 of \SI{5.68}{\meter}, showing that robust weighting contains the remaining heavy tails in BLE and WiFi once systematic anchor-mapping errors are removed.

\begin{figure}[t]
  \centering
  \includegraphics[width=\columnwidth]{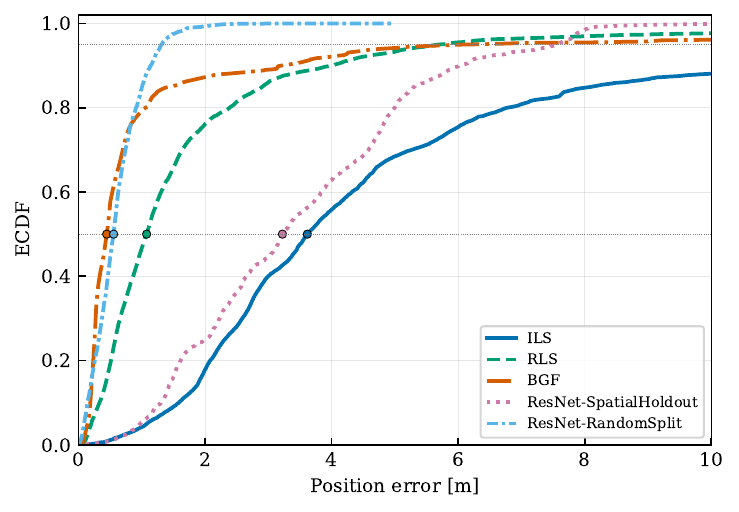}
  \caption{ECDF of horizontal positioning error on the fused input. Dotted lines mark the median and P95. The abscissa is clipped at \SI{10}{\meter}.}
  \vspace{-3mm}
  \label{fig:ecdf_fused}
\end{figure}

The error distributions in Figure~\ref{fig:ecdf_fused} show where these summary statistics diverge. BGF is tighter than ResNet-RandomSplit up to the 75th percentile. Above P75 the BGF curve flattens as grid-bounded estimates accumulate, while ResNet-RandomSplit overtakes BGF before the 95th percentile. The two ResNet variants expose the generalisation cost of the \texttt{SpatialHoldout} protocol. When reference points are held out of training, the median error increases from \SI{0.55}{\meter} to \SI{3.22}{\meter} (P95 from \SI{1.31}{\meter} to \SI{7.57}{\meter}). The \texttt{RandomSplit} number therefore represents the best-case interpolation regime and should be interpreted as such.

\begin{figure*}[t]
  \centering
  \includegraphics[width=\textwidth]{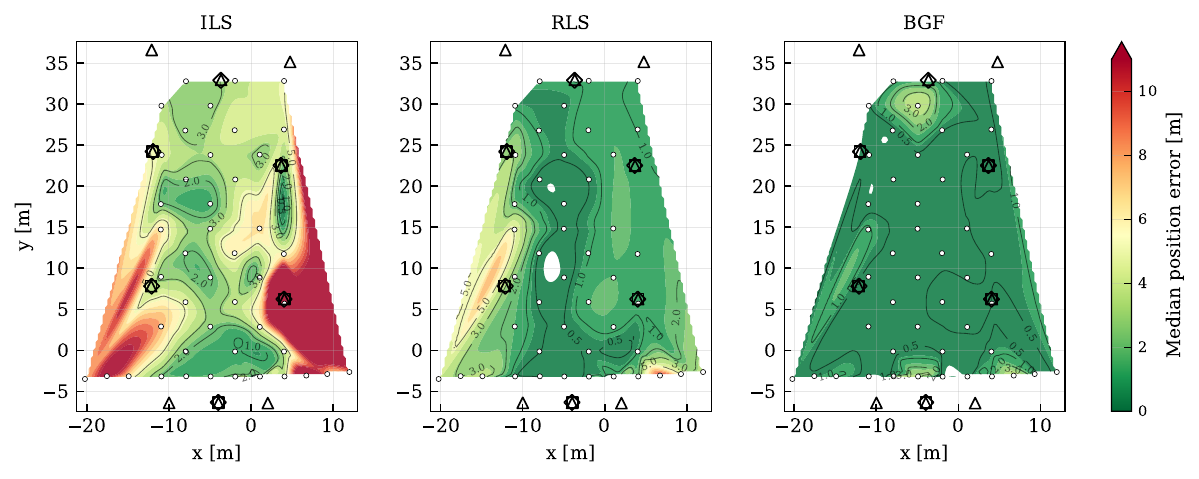}
  \caption{Interpolated median positioning error per reference point across the HYMN layout. White circles are reference points. Open squares, triangles, and diamonds are the BLE, UWB, and WiFi anchors. Panels share the right-hand colour scale.}
  \label{fig:spatial}
\end{figure*}

A spatial view for the three classical methods in Figure~\ref{fig:spatial} encodes per-reference-point median error as an interpolated heatmap over the measurement layout. The median is the consistent summary across Table~\ref{tab:positioning_stats}, Figure~\ref{fig:ecdf_fused}, and this view, so tail behaviour is left to the P95 column and the ECDF above the 75th percentile. ILS errors grow towards the room perimeter where anchor geometry degrades. RLS and BGF confine most of the residual energy to the same peripheral points but with lower absolute values. The two ResNet variants are omitted from the spatial view, since their error structure follows training-point coverage rather than anchor geometry and is already captured by Table~\ref{tab:positioning_stats} and Figure~\ref{fig:ecdf_fused}.

Two caveats qualify these numbers. The headline ResNet result applies only to the \texttt{RandomSplit} protocol, which trains and tests on overlapping reference points. The \texttt{SpatialHoldout} error, several times larger (Table~\ref{tab:positioning_stats}, Figure~\ref{fig:ecdf_fused}), is the honest indicator of generalisation. All methods share the same ranging inputs but not the same sample count $n$: the classical methods run on every valid epoch, while each ResNet protocol sees only its held-out test split. Row-wise comparison in Table~\ref{tab:positioning_stats} is therefore indicative rather than strictly paired, and the ECDFs in Figure~\ref{fig:ecdf_fused} remain the primary reference for distributional claims.

\section{Conclusion}
\label{sec:conclusion}

We benchmarked ILS, RLS, BGF, and a ResNet regressor on fused UWB, BLE, and WiFi ranges from the HYMN dataset. BGF attained the lowest median error (\SI{0.44}{\meter}) but carried a heavier upper tail (P95 \SI{6.29}{\meter}), while the ResNet under random epoch-level splitting sat close on the median (\SI{0.55}{\meter}) and dominated the tail (P95 \SI{1.31}{\meter}). RLS with Huber weighting (median \SI{1.07}{\meter}, P95 \SI{5.68}{\meter}) remained usable under BLE and WiFi noise, while ILS was outlier-dominated (median \SI{3.62}{\meter}, P95 \SI{18.66}{\meter}). Per-technology ranging quality spans roughly two orders of magnitude, which caps any fusion that treats all anchor ranges equivalently.

Holding entire reference points out of ResNet training raised median error to \SI{3.22}{\meter} and P95 to \SI{7.57}{\meter}, so the learned model's advantage under random splits reflects interpolation rather than generalisation to unseen locations. The grid filter therefore remains competitive wherever dense reference coverage cannot be assumed. Further caveats include the single-dataset scope and an unmatched sample counts between classical methods and the ResNet.

These limitations motivate cross-dataset evaluation to test whether the method ranking is stable, fusion schemes that incorporate 5G ranges and intermittent GNSS under partial line-of-sight, training protocols aimed at spatial generalisation, for instance through domain randomisation of anchor geometry, and accuracy as a function of training-set size, since survey collection is often the dominant deployment cost. Extending the benchmark to dynamic receivers would complement the static-point evaluation and expose how the ranking shifts under motion.

\section*{Data and Software Availability}
The HYMN dataset is openly available on Zenodo \cite{HYMN_Dataset_zenodo_2025} under the MIT license. A companion data descriptor \cite{Ammad2026_HYMN_descriptor} documents the acquisition setup, anchor placement, sensor models, and timestamp conventions. The evaluation code ships these measurements unchanged from the Zenodo deposit, in both CSV and pickled-DataFrame form.

The evaluation code that produced all numbers, tables, and figures in Section~\ref{sec:evaluation} is archived on Zenodo \cite{HYMN_Code_zenodo_2026} and developed at \url{https://github.com/TUD-ITVS/hymn-localization-ipin2026} under the MIT license. The repository includes a pinned \texttt{requirements.txt} and a README that walks through reproduction steps, hardware requirements, and indicative execution times. It also provides an explicit mapping from each script to the manuscript figures and tables it generates. The released code reproduces preprocessing, the four method implementations, the two evaluation protocols, and every statistic, table, and figure in Section~\ref{sec:evaluation}. Trained ResNet checkpoints are not redistributed. They regenerate from the bundled measurements and the hyperparameters in Section~\ref{sec:methodology}.

%\section*{Acknowledgment}
%The authors thank Albrecht Michler, Jonas Ninnemann, and Hagen Ußler, co-authors of the HYMN dataset, for their contributions to the dataset collection, ground truth acquisition, and measurement campaigns on which this evaluation is built.

\section*{Declaration on Generative AI}
During the preparation of this work, the authors used Anthropic's Claude (via the Claude Code CLI) to assist with drafting and editing of the manuscript as well as with development of the evaluation code. All AI-generated content was reviewed, verified, and edited by the authors, who take full responsibility for the final content of this publication.

\balance
\bibliographystyle{IEEEtran}
\bibliography{bib/references}

\end{document}